\documentclass[letterpaper]{article} 
\usepackage{aaai23}  
\usepackage{times}  
\usepackage{helvet}  
\usepackage{courier}  
\usepackage[hyphens]{url}  
\usepackage{graphicx} 
\usepackage{natbib}  
\usepackage{caption} 
\usepackage{algorithm}
\usepackage{algorithmic}

\usepackage{newfloat}
\usepackage{listings}
\DeclareCaptionStyle{ruled}{labelfont=normalfont,labelsep=colon,strut=off} 
\floatstyle{ruled}
\newfloat{listing}{tb}{lst}{}
\floatname{listing}{Listing}
\usepackage{enumitem}
\setlist[itemize]{noitemsep, nolistsep}
\usepackage{multirow}
\usepackage{booktabs}
\usepackage{csquotes}
\usepackage{amsmath}
\usepackage{makecell}
\usepackage{tcolorbox}

\title{Language Models are Drummers:\\Drum Composition with Natural Language Pre-Training}
\author{
    Li Zhang,
    Chris Callison-Burch
}
\affiliations{
    University of Pennsylvania\\

    \{zharry,ccb\}@seas.upenn.edu
}

\usepackage{bibentry}

\begin{document}

\maketitle

\begin{abstract}
Automatic music generation with artificial intelligence typically requires a large amount of data which is hard to obtain for many less common genres and musical instruments. To tackle this issue, we present ongoing work and preliminary findings on the possibility for deep models to transfer knowledge from language to music, by finetuning large language models pre-trained on a massive text corpus on only hundreds of MIDI files of drum performances. We show that by doing so, one of the largest, state-of-the-art models (GPT3) is capable of generating reasonable drum grooves, while models that are not pre-trained (Transformer) shows no such ability beyond naive repetition. Evaluating generated music is a challenging task, more so is evaluating drum grooves with little precedence in literature. Hence, we propose a tailored structural evaluation method and analyze drum grooves produced by GPT3 compared to those played by human professionals, exposing the strengths and weaknesses of such generation by language-to-music transfer. Our findings suggest that language-to-music transfer learning with large language models is viable and promising.\footnote{Accepted to the 1st workshop on Creative AI across Modalities in AAAI 2023.}\footnote{Data and code can be found at \url{https://github.com/zharry29/drums-with-llm}. The title is a parody of the viral trend of titling papers as ``Language Models are ...'' in NLP venues.}
\end{abstract}

\section{Introduction}
Music understanding and generation using artificial intelligence has a long history \cite{roads1985research} and has gained steady interests in recent years \cite{kaliakatsos2020artificial}. One strand of work focuses on symbolic music rather than audio, where music is represented as sequential data such as MIDI. While the analogy between music and language has long been studied \cite{10.1525/mp.2004.21.3.289}, the symbolic representation of music exhibits an even clearer similarity to language in their surface form. For example, music has notes, measures, and sections, while language has tokens, sentences, and paragraphs. It is thus intuitive that some work has applied natural language processing (NLP) techniques to music. Specifically, most have attempted to learn an embedding space of music \cite{liang2020pirhdy} similar to that of texts. 

\begin{figure}[t!]
    \centering
    \includegraphics[scale=0.37]{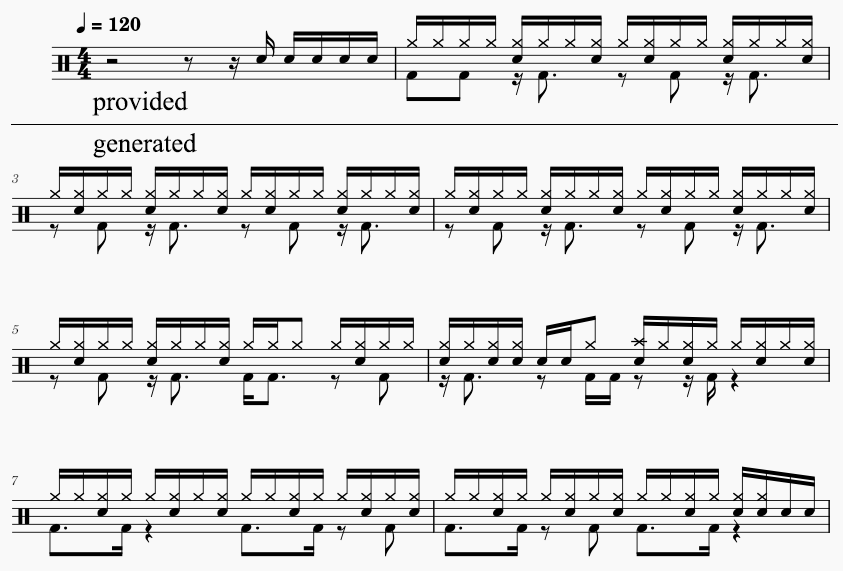}
    \caption{The sheet music of an example of our generated drum track, demonstrating our model's ability to somewhat musically follow the motif and make variations. The first two measures are provided while the rest are generated.}
    \label{fig:example}
\end{figure}

Recent work has leveraged Transformers \cite{NIPS2017_3f5ee243} for symbolic music processing \citep{huang2018music,https://doi.org/10.48550/arxiv.2210.10349}. Transformers, when pre-trained on a massive text corpus, are known as large language models (LLMs), the current state-of-the-art approach to many NLP tasks \cite{devlin-etal-2019-bert,NEURIPS2020_1457c0d6}. Notably, \citet{zeng2021musicbert} is one of the first and only work to pre-train a Transformer on a large symbolic music corpus containing more than 1 million songs, using a similar approach of pre-training on a text corpus, achieving state-of-the-art performance in various music understanding tasks. However, one of the most significant limitations of this and most other work in data-driven music processing is the large amount of symbolic music data required for training, which is extremely challenging to obtain for less-mainstream genres, particular styles, less-prevalent instruments, or out-of-the-ordinary specifications, severely limiting the versatility of such a method. Hence, low-resource symbolic music processing remains highly challenging. For texts, on the other hand, few-shot learning has been greatly empowered by LLMs due to the extremely large size of their pre-training textual data, magnitudes more than music data.


In this work, we are the first step to explore such \textbf{text-to-music transfer learning potential in LLMs}. In other words, we pose the hypothesis that present-day state-of-the-art LLMs, pre-trained with a massive amount of textual data, is capable of \textbf{generating symbolic music with little music data} to some nontrivial extent. We specifically focus on one instrument, the drum set, for multiple reasons. First, the drum set is one of the most common and important instruments in many genres of music such as jazz, funk, blues, gospel, latin, pop, rock, metal, etc. Second, the symbolic representation of the drum set is simpler than most pitched instruments, as each note does not have a pitch but corresponds to a hit on one drum. As the number of drums is usually greatly smaller than that of possible pitches, the resulting sequence is much shorter, and thus easier to be processed by models. Third, the performance of a drum set typically is endowed with more degree of freedom with regard to the audience's aesthetics than many other instruments, making it an appropriate entry point for studying music generation with LLMs, which is presumed to be highly challenging. 

We focus on the task of drum generation or composition, which has a small body of published work in literature. While most if not all all of the existing work has treated drums as an accompaniment, we instead focus on drum solo generation, with a convenient analogy to story generation at which LLMs are known to excel.

We finetune a state-of-the-art LLM, GPT3 model \cite{NEURIPS2020_1457c0d6} on the Groove dataset \cite{gillick2019learning} of about 400 drum groove performances recorded as MIDI. To leverage the textual pre-training of GPT3, we propose a textual representation of a drum performance. We present the following core findings:
\begin{enumerate}
    \item The largest and smallest GPT3 models both can generate nontrivial drum grooves after being finetuned.
    \item A similar-sized model that is not pre-trained on language data, however, cannot.
\end{enumerate}
We claim that the existing automatic evaluation of music generation is insufficient for our task. Hence, we propose an evaluation methodology specifically for drum grooves to both qualitatively and quantitatively evaluate the strengths and weaknesses of machine-generated drum grooves compared to those performed by humans. Finally, we provide some preliminary listening test results, with a plan to conduct scaled and rigorous tests in future work.

\section{The Drum Set}
\begin{figure}[t!]
    \centering
    \includegraphics[scale=0.53]{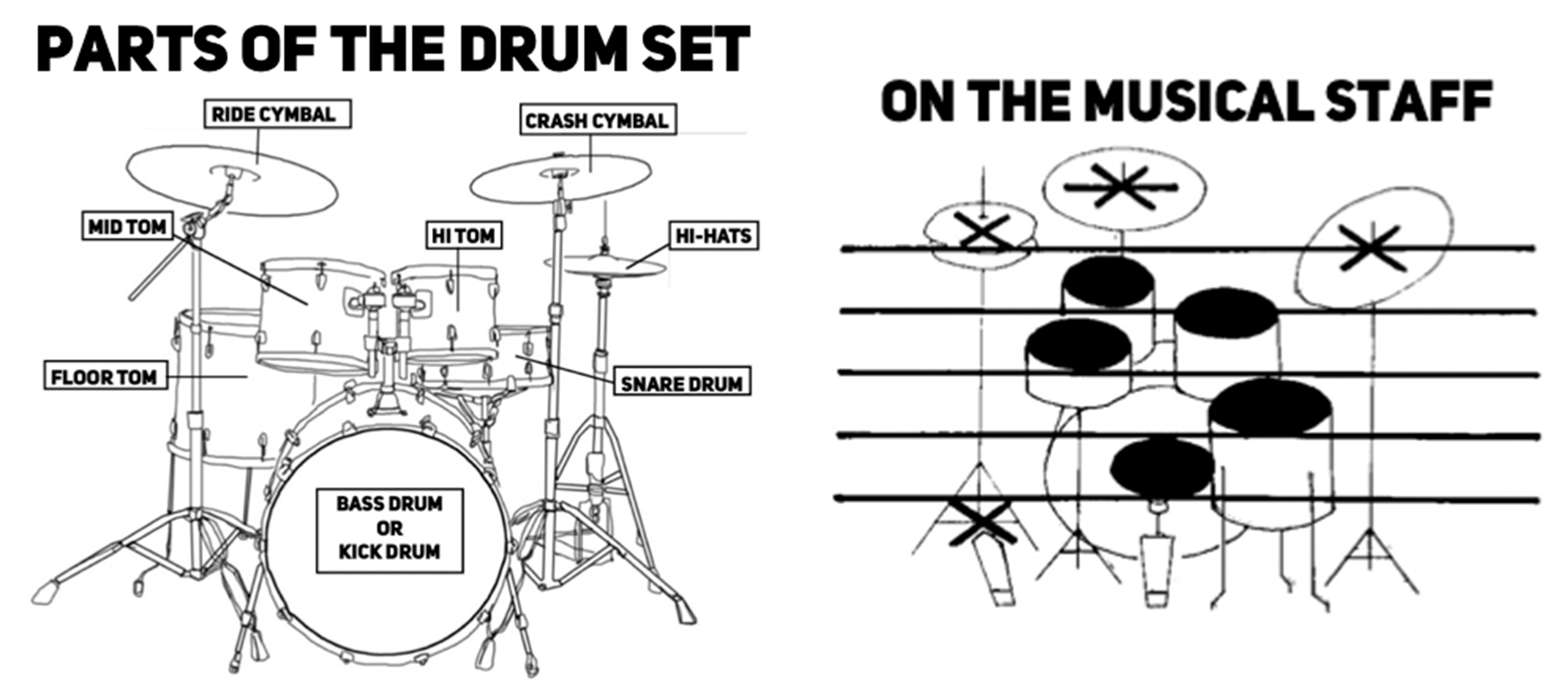}
    \caption{The anatomy of a basic drum set, and how each drum may appear on a sheet music.}
    \label{fig:drumset}
\end{figure}

The drum set, also known as the drum kit, or colloquially the drums, is a compound musical instrument consisting of many sub-instruments, including drums and cymbals, both generally referred to as drums in this paper. Here, we assume a simple, stereotypical drum set with a hi-hat, a crash cymbal, a ride cymbal, a bass drum, a snare drum, and a tom (see Figure~\ref{fig:drumset}, but without the hi-tom and the floor-tom\footnote{\url{https://i.redd.it/cen72phsihb81.png}}). 

The performance on a drum set can be notated as sheet music, a human-readable symbolic representation, or as MIDI, which records the when each drum is hit at what velocity, a computer-readable symbolic representation. 

\section{Dataset}
\begin{table}[t!]
\centering
\begin{tabular}{llll}
\toprule
            & train & dev & test \\ \midrule
Total num. MIDI   & 373   & 47  & 35   \\
- rock &   169    &  16   &  15    \\
- jazz &   41    &  6   &   4   \\
- latin &   37    &  10   &  3    \\
- funk &    31   &  6   &   4   \\
- hiphop &   26    &  1   &    3  \\
- others &   69    &  8   &   6   \\
\bottomrule
\end{tabular}
\caption{The number of MIDI files in each style in the filtered Groove dataset used in this work.}
\label{fig:dataset_stats}
\end{table}

Among just a few datasets of drum performances, Google's Groove MIDI Dataset\footnote{https://magenta.tensorflow.org/datasets/groove} is the largest and the most high-quality to date, containing 1,150 MIDI files and over 22,000 measures of drumming by 10 professional drummers. In this dataset, the drum performances are either grooves, long sequences of rhythmic ideas, or fills, short bursts of free-flowing expressions. As we focus on drum generation or composition throughout a long sequence, we only consider the grooves in the dataset. Each MIDI is marked with the style (e.g., rock, funk, gospel, etc.), the tempo (in beats per minute, BPM), and the time signature. For simplicity, we only consider those in the time signature of 4/4. We follow the train-development-test splits in the dataset. The statistics of the filtered Groove dataset are shown in Figure~\ref{fig:dataset_stats}. 

The Groove dataset was originally proposed to study microtiming and expressive performance, and therefore the drum MIDI files encapsulate human imperfection. However, we re-purpose the dataset to study drum composition, leading to the following choice of simplification. First, we quantize all notes to a 16-th note grid. In other words, all note events in the MIDI are re-timed to the closest of the 16 equidistant timestamps in a measure. An implication of such quantization is that deliberate off-grid playing such as triplets or swing feels is lost. Second, we discard the velocity information (i.e., how hard a drum is hit), which can usually be inferred post-hoc to a coarse-grained extent. Third, while the drum set can be played with many expressions and articulations (e.g., hitting different part of a drum using different part of different tools), we reduce them to simply the basic articulation (head hit) of the hi-hat, crash cymbal, ride cymbal, bass drum, snare drum, and floor tom. In other words, each note only has 6 possible values. Fourth, we truncate each MIDI file to only the first 16 measures; at 128 BPM, for example, this equates to 30 seconds. Finally, we remove empty leading measures whose first quarter note is a rest, and ignore grooves with less than 8 measures.

\section{Experiments}

\subsection{Representation}

\begin{figure}[t!]
    \centering
    \includegraphics[scale=0.46]{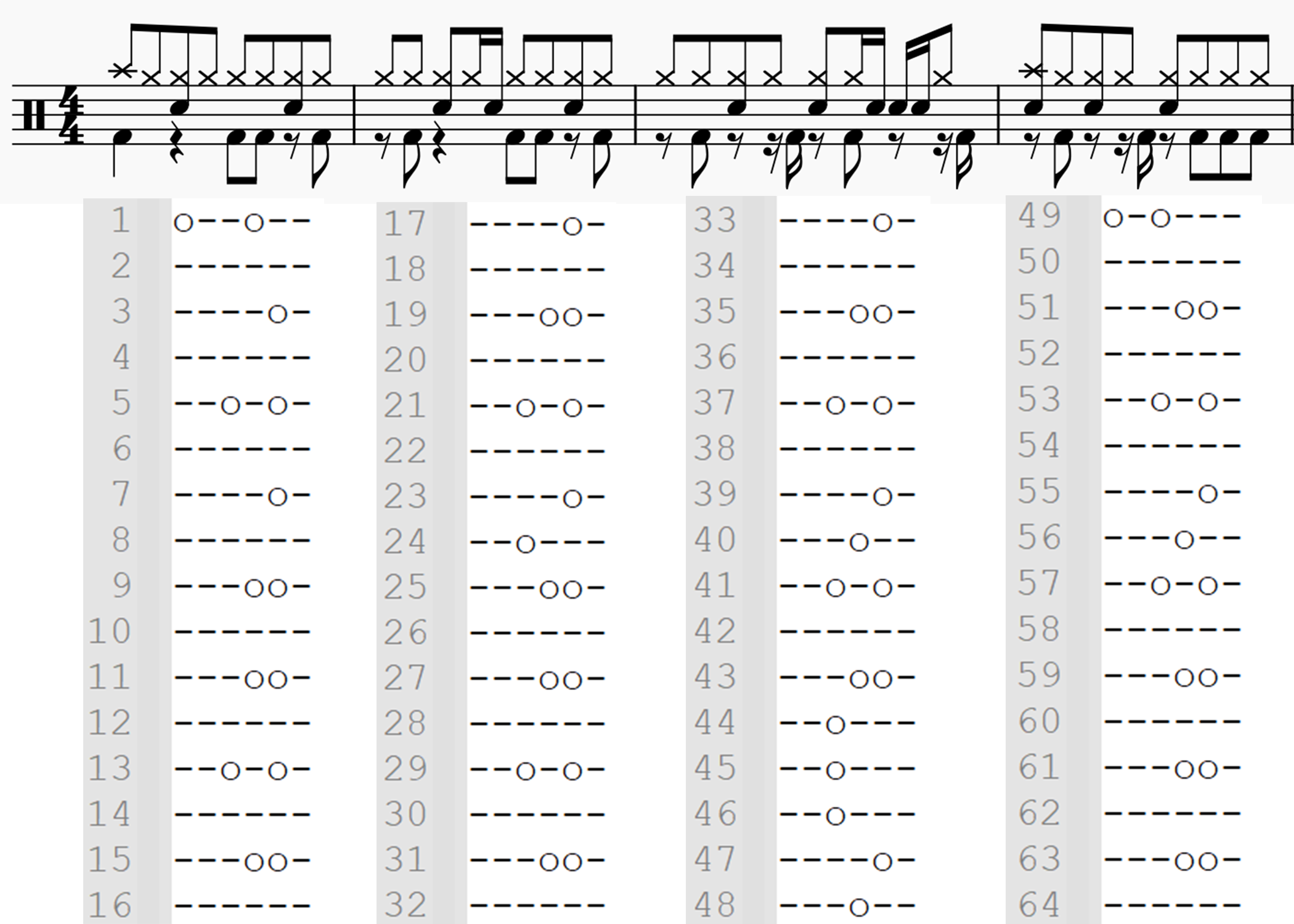}
    \caption{The drumroll representation of an example drum sheet music. Each measure corresponds to 16 lines in texts, where each line corresponds to a 16-th note. Each line contains 6 characters corresponding to 6 drums, where `o' and `-' denotes whether each drum is hit.}
    \label{fig:drumroll}
\end{figure}

As discussed before, LLMs have demonstrated extremely strong few-shot transfer learning ability from one textual task to other. To possibly exploit this in music, it is necessary to come up with a textual representation of the drum grooves. We propose a pianoroll-like representation \cite{brunner2018symbolic}, referred to as a \textit{drumroll}, that is essentially a multi-line string where each row corresponds to a 16-th note in the time sequence, and each character in a line corresponds to whether a drum is hit. Specifically, each character is `o' if the particular drum is hit at the particular 16-th note, and `-' otherwise. See Figure~\ref{fig:drumroll} for an example. To help LLMs identify the boundary between measures, we add a newline of ``SEP'' between every 16 lines (a measure) and a newline of ``END'' after the final line.

\subsection{Task}
We focus on an instance of drum generation referred to as drum completion, where the model is given the first 2 measures and must complete the rest of the 14 measures of the groove. The model may terminate at any point. This is analogous to conditioned story generation in NLP. 

\subsection{Model}
First, we consider two naive baselines, \textbf{random}ly choosing whether to play a note and \textbf{repeat}ing the second given measure. We then finetune a state-of-the-art LLM, OpenAI's \textbf{GPT3 Davinci} with 175 billion parameters, on the training set. In each file of a drum groove, the input (prompt) is the first 2 measures and the output (completion) is the remaining 14 measures. The temperature is set to 0.85 to encourage creativity. Finetuning the model on the training set costs \$38.33 and takes around 30 minutes using OpenAI's API.

To ascertain the role model size plays in drum generation, we further consider a smaller \textbf{GPT3 Ada} model with 350 million parameters, which has been pre-trained on the same corpus and we use the identical settings as the larger Davinci model. Finetuning the model on the training set costs \$0.51 and takes around 5 minutes using OpenAI's API. Both GPT3 models are later found to be able to generate nontrivial drum grooves.

The ascertain the role language pre-training plays in drum generation, we set the control to be an un-pre-trained GPT3 model, namely a Transformer \cite{NIPS2017_3f5ee243} with the same size as GPT3. Because our computing resources cannot accommodate a model as big as Davinci, and also because merely 373 text files each with 256 lines in the training set are likely insufficient to converge the training loss, we finetune a smaller, un-pre-trained Transformer with 85 million parameters, the same magnitude as GPT3 Ada. While the training loss does converge, the model predicts the same certain sequence regardless of what 2 measures are provided, performing no better than the random baseline. This suggests that language pre-training is a necessary condition for effective drum generation.
\begin{figure*}[t!]
    \centering
 \includegraphics[scale=0.55]{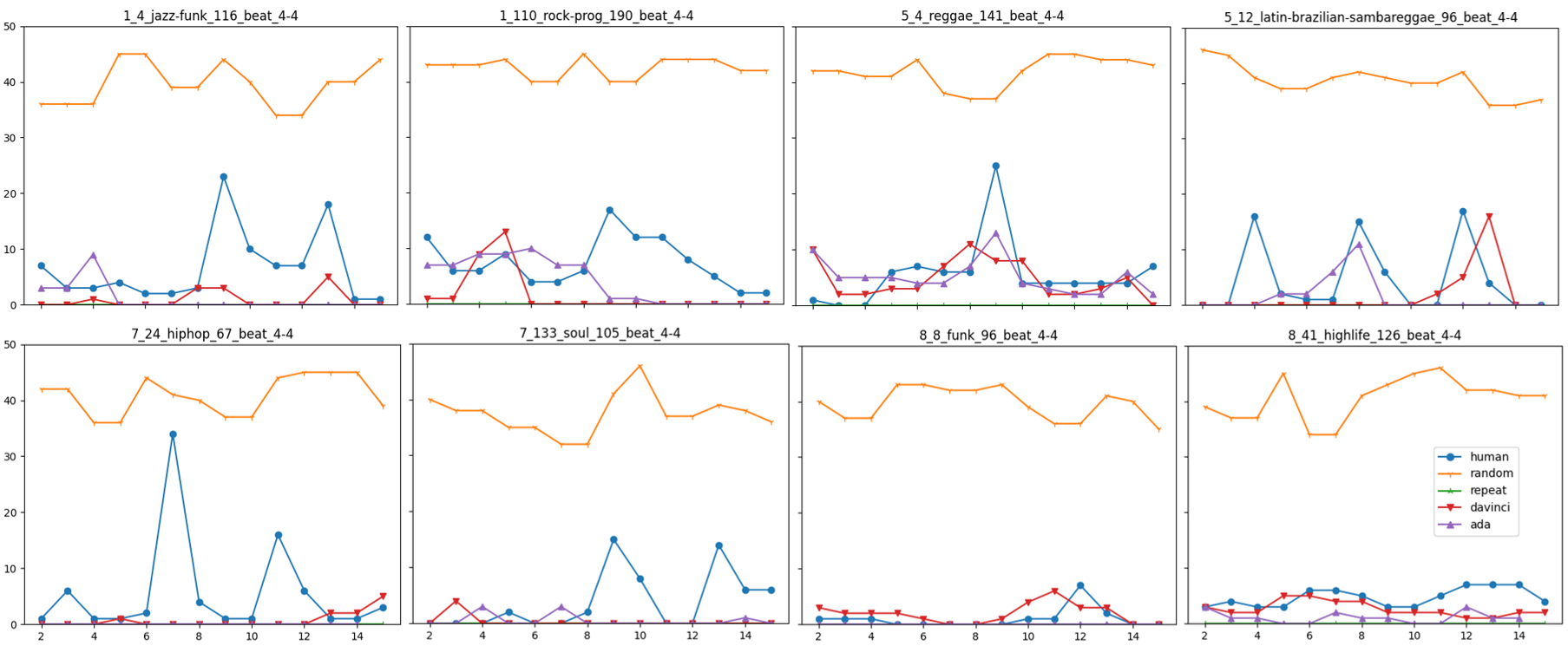}
    \caption{The variation (y-axis) of each measure (x-axis) in 8 randomly sampled drum grooves from each style.}
    \label{fig:repetition_variation}
\end{figure*}
\section{Evaluation}
How good are the drum grooves generated by GPT3? We follow the convention of music generation and consider both objective/automatic and subjective/human evaluation. We report our findings based on the test set.

\subsection{Objective Evaluation} 

We deem the established methods of automatic evaluation of symbolic music generation unsuitable for our task. To name a few: perplexity, for example, has long been shown to exhibit low correlation with human perception in NLP \cite{kuribayashi-etal-2021-lower}; structural similarity, while a reasonable metric, is often approximated with some simple similarity measures \cite{https://doi.org/10.48550/arxiv.2210.10349} and discourages creativity that stray from the reference, which by no means should be treated as the ground-truth. 

In contrast, we consider what constitutes a good drum groove and propose a structural evaluation called the \textbf{pattern and fill analysis}. We assume that often a good drum groove minimally satisfies the following criteria: 
\begin{enumerate}
    \item There exists one or more consistent \textbf{patterns} of some rhythmic idea and occasional change-ups known as \textbf{fills}.
    \item The measures in a pattern are sufficiently similar, but ideally not identical.
    \item The measures in a fill are sufficiently different from those in adjacent patterns.
\end{enumerate}

Later, we verify that the human-performed grooves mostly satisfy these criteria. 

In a drum performance represented as a drumroll, each measure is represented by a string of 16 lines (Figure~\ref{fig:drumroll}). To classify each measure as either a pattern or a fill, we take a sliding window of size 3 centered at some measure $m_i$ and calculate the edit distances between this measure and its two neighbors. The minimum of these two distances is referred to as the \textit{variation} of the central measure:
$$\textit{variation}(m_i) = \min (dist(m_i,m_{i-1}),dist(m_i,m_{i+1}))$$

Intuitively, the variation of a measure in a pattern would be small, while that in a fill would be large. Therefore, in a good drum groove, the variation of all measures (except the first and the last) can be expected to be clearly separated. 

\begin{table}[t!]
\small
\centering
\begin{tabular}{llllll}
\toprule
            & human & random & repeat & Davinci & Ada \\ \midrule
avg. variation   &  5.1  &  40.4 & 0  & 3.0  &  3.4   \\
\makecell[l]{avg. intra- \\ centroids}   &  10.1  &  5.7 & 0  & 6.9  &  7.2   \\
\makecell[l]{avg. inter- \\ centroids}   &  1.4  &  1.2 & 0  & 0.6  &  0.8   \\
\bottomrule
\end{tabular}
\caption{Qualitative statistics of drum grooves produced with different means. The distance between cluster centroids 
 suggests how dissimilar the patterns and the fills are. The distance between a cluster centroid and a measure suggests the amount that the measures in one class vary by.}
\label{tab:variation}
\end{table}

Next, we plot the variation of each measure sequentially for each drum groove in the development set and observe consistent patterns. Eight randomly chosen grooves of different styles are shown in Figure~\ref{fig:repetition_variation}. Intuitively, for human performances, the variation of a measure in a pattern should be small, while that in a fill should be large. This is indeed the case for most examples, where human-played patterns are consistent but with some variations (plateaus), while fills are largely different (spikes). As expected, the random baseline results in high variation across all measures, while the repeat baseline no variation at all -- both are undesirable. 

\begin{table}[t!]
\small
\centering
\begin{tabular}{llllll}
\toprule
            & human & random & repeat & Davinci & Ada \\ \midrule
Repetitive   &  0  &  0 & 35  & 3  &  7   \\
Consistent   &  32  &  0 & 0  & 29  &  15   \\
Chaotic   &  3  &  35 & 0  & 3  &  13   \\
Has fill   &  30  &  0  & 0  &  13  &  10  \\
Avg. length   &  13.3  &   16  & 16  &  13.7  &  12.5  \\
\bottomrule
\end{tabular}
\caption{The number of drum grooves judged to satisfy each criteria produced by each model in the test set.}
\label{tab:subjective}
\end{table}

Upon qualitative examination, the two GPT3 models of different sizes can clearly generate nontrivial drum grooves, with many plateaus and occasional spikes, though less spikes than those performed by human. Quantitatively, we calculate the average variation of all measures in all grooves. As shown in Table~\ref{tab:variation}, drum grooves generated by GPT3 models tend to vary less than human. To see if the generated patterns have less variation while the fills have much more variation, we perform K-means to separate the measures in a drum groove into two clusters by their variation. We then calculate the average intra-distance between the two centroids, and the average inter-distance between each measure and the centroid it is assigned to. As shown in Table~\ref{tab:variation}, the intra-centroid distance shows that the grooves performed by human have a much clearer-cut pattern-versus-fill separation than GPT3, than the random baseline. The inter-centroid distance shows that the spread of variations within the class of pattern or fill is more pronounced in human-performed grooves than in GPT3-generated ones.

\begin{figure*}
    \centering
    \includegraphics[scale=0.63]{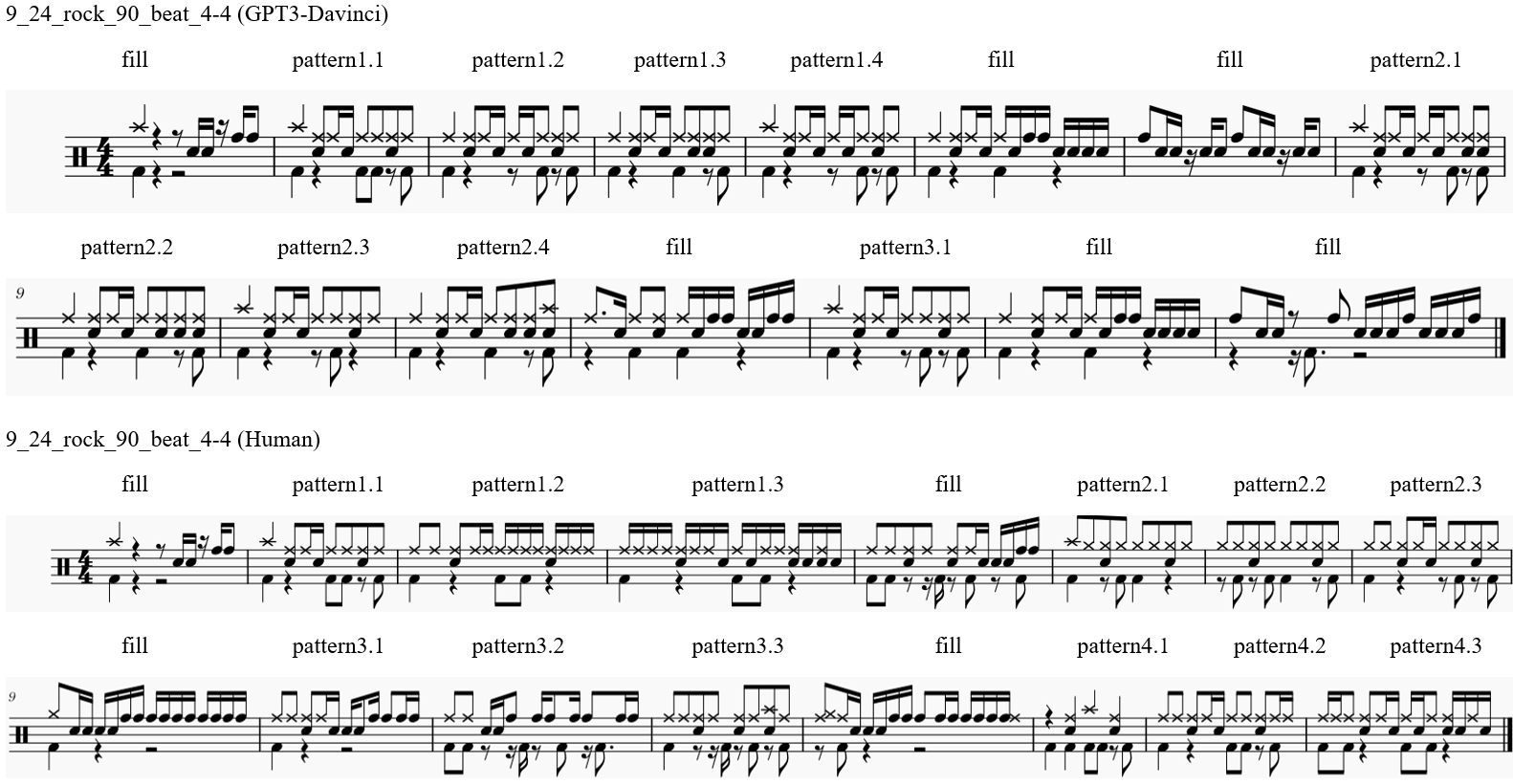}
    \caption{The sheet music of a satisfactory drum groove generated by GPT3 (above), juxtaposed with the groove with the same first 2 measures played by human from the dataset (below). We annotate each measure as either a pattern or a fill. For example, pattern$x$.$a$ and pattern$x$.$b$ are adjacent, have the same accented back-beats, but are not identical.}
    \label{fig:good_example}
\end{figure*}

\subsection{Subjective Evaluation}
Our objective evaluation is clearly informative but insufficient. Hence, we conduct a listening test and perform an error analysis based on the following criteria:
\begin{itemize}
    \item Is the groove repetitive, meaning there is little or no variation among measures?
    \item Is the groove consistent, meaning there is some variation among measures but a steady rhythmic idea (specifically, the back-beat placement) can be followed?
    \item Is the groove chaotic, meaning there is either too much variation, or a lack of a clear rhythmic idea?
    \item Does the groove contain any reasonable drum fill?
\end{itemize}

While scaling up this analysis rigorously with carefully chosen subjects is left for future work, our own judgements are shown in Table~\ref{tab:variation} as preliminary findings. Concretely, all drum grooves produced via different means are shuffled and randomly present to one of the authors who has had years of training in drumming. Naturally, all randomly generated grooves are judged as chaotic without any consistent motif, while all repeated ones are by nature repetitive. For the grooves performed by human, most are judged as consistent and most include at least one fill which is sufficiently different from the rest of the rhythmic patterns. In comparison, grooves generated by GPT3 Davinci is only slightly less consistent with desirable variations among measures, but significantly less of them contain any fills, rendering the grooves less interesting and more predictable overall. For the smaller GPT3 Ada model, the observation holds to a larger extent, with more inconsistent grooves and less fills. 

\begin{figure*}
    \centering
    \includegraphics[scale=0.48]{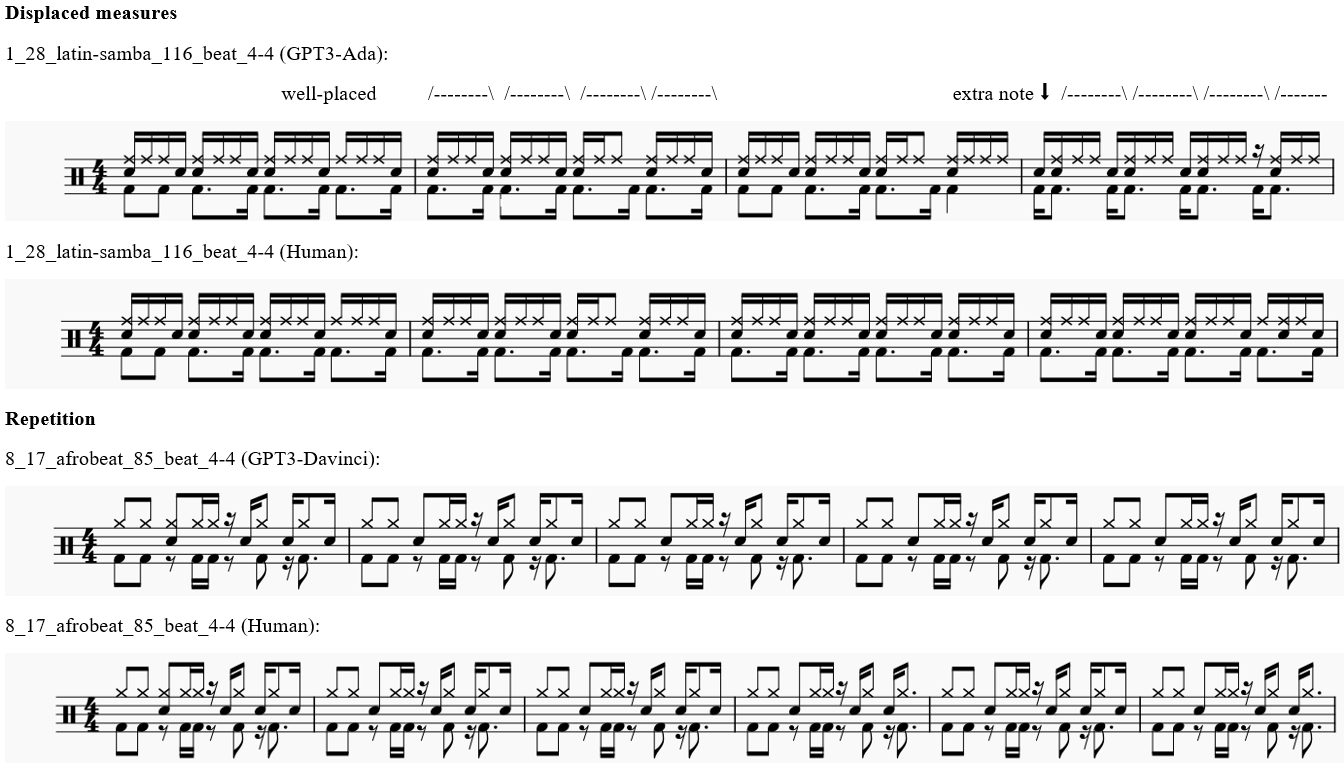}
    \caption{The sheet music of two problematic drum grooves generated by GPT3 (the 1st and 3rd), juxtaposed with those performed by human (the 2nd and 4th). In the first case, GPT3 generates an extra note at the beginning of the fourth measure, displacing all grooves afterwards. In the second case, GPT3 generates strictly repeated measures; the rest are omitted.}
    \label{fig:bad_example}
\end{figure*}

\subsection{Case Study and Takeaways}
We claim that the LLMs demonstrate an impressive ability to write good drum grooves given only hundreds of training examples without any knowledge of what features to pay attention to. From our objective and subjective evaluation above, we have observed that both LLMs and human are able to compose drum grooves with a structure and some variations. Why, however, are models' drum composition still worse than human's? We postulate several factors.

We observe that a common source of the lack of musicality in machine-generated drum grooves stems from the misplacement of back-beats, which are steadily accented beats in a measure, usually the 2nd and the 4th in 4/4 in many genres of music. Human drummers, when playing variations, tend to respect the back-beat placements, while models tend to disregard such concept. Another source of the lack of musicality is the lack of drum fills. While human drummers often inherently think about the cycle between patterns and fills to give a captivating performance, models are yet to realize the importance of those occasional deviations. 

To qualitatively examine these claims, we showcase two sets drum grooves composed by GPT3, one showcasing its strengths and one its weaknesses, along with the human performed grooves with the same first 2 measures. 

A positive example is shown in Figure~\ref{fig:good_example}. Clearly, both grooves produced by human and GPT3 alike contain consistent patterns interspersed with fills, a prototypical structure that is expected and desired by most audience. Closely examining the pattern groupings, both grooves vary the placement of the bass drum and the snare drum, known as syncopation, while keeping the back-beats at the 2nd and 4th quarter note in tact. As a typical drum groove, patterns in both grooves consistently place a snare drum hit in all patterns (with two exceptions in human's pattern 3.1 and 3.2). Such composition is highly desirable, as the steady back-beat placement guides the listener with a solid rhythmic foundation, while the variations keep the groove interesting and natural. Besides the ability to conforming to human drummers' standard practice, this example also showcases LLMs' ability to be creative. While the human-performed groove closely adheres to the template of 3 measures of patterns followed by 1 measure of fill, the GPT3-generated groove does not. Instead, it employs multi-measure fills, and more interestingly, grouping of an odd number of patterns followed by fills, which is uncommon in the genre.

Two negative examples are shown in Figure~\ref{fig:bad_example}, representing two common problems of GPT3-generated drum grooves. This first case is referred to as \textit{displaced measures}. When human drummers play a pattern, they often fit each measure with some rhythmic idea. In other words, the boundaries between two measures are often clear. In our drumroll representation, each measure, 16 lines of texts, is followed by a newline of `SEP' to denote such boundary. However, GPT3 Ada sometimes fail to respect these boundaries and displaces the measures, thus having more ``chaotic'' patterns as reported in Table~\ref{tab:subjective}. In comparison, GPT3 Davinci makes a lot less such mistakes. The second case demonstrates LLMs' known drawback of the proclivity to repeat generations. In the given example, GPT3 Davinci generates 14 identical measures, while a human drummer would ``sneak in'' minor variations while maintaining the motif. It is worth noting that the repetition of drum grooves is accepted in some music genres such as pop, dance, or rock, but frowned upon in most others as they are often thought to be mechanical. For drum grooves generated by LLMs, repetition is logically a trait to be avoided.

As of now, we have taken a model-free approach where the LLMs are only trained on some drum groove data without any additional information or priors. To reinforce the strengths and alleviate the weaknesses discussed before, we suggest future work to take a modular approach instead of tackling the task in an end-to-end fashion. For example, the patterns and fills can be generated from different distributions or by different models; repetition can be explicitly discouraged by over-generating each measure and perform some voting or selection. Furthermore, it is possible to control the variation within a drum groove by injecting additional labels in the training data to condition the LLM on. 

\subsection{Improvisation or Recitation?}
By far, we have hinted at LLMs' ability to be creative with regard to drum composition. Here, we pose one additional question: are the LLMs really \textit{creating} some drum grooves that they have never seen during finetuning, or are they simply \textit{regurgitating} what they have already seen?

\begin{figure}
    \centering
    \includegraphics[scale=0.6]{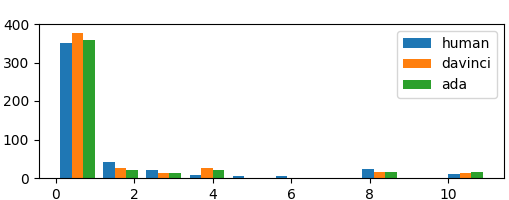}
    \caption{The count (y-axis) of occurrences (x-axis) of each generated measure appearing in the training set.}
    \label{fig:duplication}
\end{figure}

We answer this question by calculating the frequency of each generated measure in the test set appearing in the the training set. As shown in Figure~\ref{fig:duplication}, only a small portion of generated measures are duplicates of any seen measures during finetuning, suggesting LLMs' ability to compose novel and unseen drum grooves.

\section{Related Work}

\textbf{Automatic music generation} has a long history, and recent work has focused on using artificial intelligence \cite{kaliakatsos2020artificial} and specifically deep neural networks. Such efforts are driven by not only the prospect of having AI aid or replace human composers and arrangers for a variety of music, but also the pursuit of probing the artistic creativity of the state-of-the-art data-driven models. Most modern work in automatic music generation leverages model architectures that have shown to be effective in computer vision or NLP, such as LSTM \cite{lyu2015modelling}, Transformers \citep{huang2018music,zeng2021musicbert,https://doi.org/10.48550/arxiv.2210.10349}, or custom architectures to learn an embedding space \cite{liang2020pirhdy}. However, the majority of the work on music generation with AI has happened in a supervised setting, which greatly limits its application to lesser known genres or specific instruments, such as drum generation. While there is yet to be a pre-trained music generation model as versatile as its counterpart in NLP such as GPT3, we believe the exploration of language-to-music transfer is necessary but uncharted.

\textbf{Automatic drum generation} has a small body of existing work. Similar to music generation in general, the drum generation task can happen in various settings. In the simplest setting, only one measure of drum pattern (also known as a ``beat'') is generated that is supposed to repeat throughout a song  \cite{vogl2017intelligent,bruford2020jaki,complexis21}, while we focus on a more involved setting of non-repetitive drum composition. Alternatively, some work simply considers the general rhythm \cite{lattner2019high}, but not the orchestration of different drums that we emphasize on. In more practical settings, a long sequence of drum composition is generated conditioned on musical signals such as the basslines \cite{makris2017combining,makris2019conditional}. While this line of work is most similar to ours which by far only deal with drum solo performance, all the work above has used drum composition data with limited size and variations as well as models (such as LSTM) that are relatively outdated in the AI community. Nevertheless, a direct comparison would be beneficial in future work. Less related is another line of work focuses on the microtiming and humanization of drum performance \cite{gillick2019learning,burloiu2020adaptive,burloiu2020interactive}, striving to mimic human's expressive imperfections.

A small body of work has focused on rhythm games \cite{donahue2017dance,liang2019procedural}. While rhythm games and drumming have certain similarities such as the focus on note placement with regard to the rhythm, their core difference lies in that the choreography of rhythm games is optimized for difficulty and playability, while the composition of drums is optimized for musicality. Due to this fundamental difference of motivation, we consider this line of work to be mostly irrelevant to ours.

\textbf{Large language models} (LLMs) are deep neural models, such as Transformers, pre-trained on a massive text corpus. For example, GPT3 is pre-trained on the compilation of Common Crawl, containing most texts in the world wide net, publicly available books, Wikipedia, and so on. From BERT \cite{devlin-etal-2019-bert} to GPT3 \cite{NEURIPS2020_1457c0d6}, LLMs have been dominant is most tasks and applications in NLP. While much about how LLMs work is unknown, and thus LLMs are notoriously known as black-boxes, there is a generally consensus that LLMs' power can be attribute to the large size of both the pre-training data and the model, which give rise to LLMs' ability to effectively adapt to low-resource domains via transfer learning by being finetuned on a small amount of data. Interestingly, a few recent work has found that some of these abilities include transfer from pre-trained language to non-language tasks, such as chess \cite{stockl-2021-watching}. While non-language textual representations such as chess moves or music charts are similar to natural language superficially, each manifests vastly different structures, and so we claim that transfer learning between them is well worth studying. 

\section{Conclusion and Future Work}
Our preliminary findings show that pre-trained large language models (LLMs) finetuned with merely hundreds of symbolic music files, such as drum grooves, can learn to generate music non-trivially. We also provide evidence that such ability can be attributed both the model size and the presence of language pre-training. We hope this observation inspires research efforts in not only low-resource music generation, but also exploration of extraordinary potentials of LLMs. We also attempt to pioneer the automatic evaluation of drum grooves which we hope to facilitate future work in drum generation with AI. 

We also briefly discuss our plans for future work. While drum generation is scarce in literature and challenging to reproduce, we will still strive compare some existing specialized models. While our current drumroll representation ignores velocity, such information can easily be encoded by replacing the marker `o' with the velocity value. However, the effects of doing so remains to be explored. Our methodology might be ported to other instruments such as piano, which we plan to explore, whereas it would be more involved to tackle multi-instrument conditioned generation.

\newpage
\bibliography{custom,anthology}

\end{document}